\documentclass[11pt]{article}

\makeatletter
\@addtoreset{equation}{section}
\makeatother

\usepackage{epsfig,amsfonts}
\usepackage{amsmath}
\usepackage{amsthm,amssymb}
\usepackage{yfonts,mathrsfs}
\usepackage{graphicx}
\usepackage{hhline}
\usepackage{cite}
\usepackage{upgreek}
\def\be{\begin{equation}}
\def\ee{\end{equation}}
\def\bal{\begin{align}}
\def\eal{\end{align}}
\def\bea{\begin{eqnarray}}
\def\eea{\end{eqnarray}}

\def\be{\begin{equation}}
\def\ee{\end{equation}}
\def\bdm{\begin{displaymath}}
\def\edm{\end{displaymath}}
\def\bea{\begin{eqnarray}}
\def\eea{\end{eqnarray}}

\def\XXint#1#2#3{{\setbox0=\hbox{$#1{#2#3}{\int}$}
    \vcenter{\hbox{$#2#3$}}\kern-.5\wd0}}

\def\tx{\text{x}}
\def\ty{\text{y}}
\def\tt{\text{t}}
\def\tz{\text{z}}
\def\tzb{\bar{\rm z}}
\def\ptz{\partial_{\rm z}}
\def\ptzb{\partial_{\bar{\rm z}}}

\begin{document}

\begin{titlepage}

\begin{flushright}
RUNHETC-2011-01\\
\end{flushright}

\vspace{1.5cm}

\begin{center}
\begin{LARGE}
{\bf Inelastic scattering and elastic amplitude\\
in Ising field theory in a weak magnetic field\\
 at $T>T_c$. Perturbative analysis.}

\vspace{0.3cm}

\end{LARGE}

\vspace{1.3cm}

\begin{large}

{\bf A.~Zamolodchikov}$^{1,2}$, {\bf I.~Ziyatdinov$^{1}$}

\end{large}

\vspace{1.cm}

{${}^{1}$NHETC, Department of Physics and Astronomy\\
     Rutgers University\\
     Piscataway, NJ 08855-0849, USA\\

\vspace{.2cm}

${}^{2}$L.D.~Landau Institute for Theoretical Physics\\
  Chernogolovka, 142432, Russia
}

\vspace{1.5cm}

\centerline{\bf Abstract} \vspace{.8cm}
\parbox{11cm}{
Two-particle scattering in Ising field theory in a weak magnetic field
$h$ is studied in the regime $T>T_c$, using perturbation theory in $h^2$.
We calculate explicitly the cross-section of the process $2\to 3$ to the
order $h^2$. To this order, $\sigma_{2\to 3}$ dominates
$\sigma_\text{tot}$ (the probability of all inelastic processes) at all energies $E$.
We show that at high energy the $h^2$ term in $\sigma_\text{tot}$ grows as
$16 G_3 \,h^2\,\log E$, where $G_3$ is exactly the third moment of the
Euclidean spin-spin correlation function. Going beyond the leading order, we argue
that at small $h^2$ the probability of the $2\to 2$ process decays as
$E^{-16 G_3\,h^2}$ as $E\to \infty$.
}
\end{center}

\bigskip
\bigskip

\begin{flushleft}
\rule{4.1 in}{.007 in}\\
{February  2011}
\end{flushleft}
\vfill

\end{titlepage}

\newpage


\section{Introduction}

The Ising Field Theory (IFT) is the scaling limit of the 2D Ising model in a magnetic
field $H$, taken near its critical point $T=T_c$, $H=0$. In field-theoretic terms,
it can be defined as the unitary conformal field theory with $c=1/2$, perturbed by
its two relevant operators,
\be\label{ift}
{\cal A}_{\rm IFT} = {\cal A}_{\rm c={1/2}\ CFT}\, -\,
\frac{m}{2\pi}\,\int\varepsilon (x)\,d^2 x\, + \,h \int\sigma(x)\,d^2 x\ .
\ee
The field $\varepsilon(x)$, of conformal dimensions $(1/2,1/2)$, represents
the temperature deviation from the critical point (i.e. $m \sim (T-T_c)$); it
is usually called the ``energy density''. The ``spin density'' $\sigma(x)$
has the dimensions $(1/16,1/16)$, and the associated coupling parameter $h$
is the rescaled magnetic field\footnote{We assume the standard CFT normalization
of these fields:
\[
\langle \varepsilon(x)\varepsilon(0)\rangle \to |x|^{-2}\,,\quad
\langle \sigma(x)\sigma(0)\rangle \to |x|^{-1/4}\quad \text{as} \quad |x|\to 0\,.
\]
Also, note that the sign of $m$ is changed as compared to the notations in \cite{fz1,fz2}.}.
Away from the critical point $m=0$, $h=0$ the theory
\eqref{ift} is massive\footnote{Exception is the Yang-Lee point, which at real positive
$m$ appears at special purely imaginary $h=\pm\,i\,m^{15/8}\,(0.18933(2))$. These
are critical points \cite{Fisher}, where the correlation length diverges, i.e. the mass
of the lightest particle measured in the units of $m$ vanishes. The associated CFT is $c=-22/5$
non-unitary Minimal model \cite{Cardy}.}, and its physical content can
be understood in terms of the spectrum of particles and their scattering amplitudes.
The parameters $m$ and $h$ carry mass dimensions
$1$ and $15/8$, respectively. Thus, apart from the overall mass scale, the theory
\eqref{ift} depends on a single dimensionless parameter
$\eta = m/|h|^{8/15}\,$.

The theory \eqref{ift} admits an exact solution at two special (``integrable'') points. One is
the case of zero magnetic field, $h=0$, where \eqref{ift} reduces to a theory of free
Majorana fermions with mass $|m|$ (see e.g. \cite{itzykson}). This corresponds to the
values $\eta = \pm\,\infty$, where the sign of the infinity distinguishes between
two phases of the $h=0$ theory. At $\eta = +\infty$ the spin-reversal symmetry
is spontaneously broken, and the field $\sigma(x)$ develops a nonzero expectation
value $\langle \sigma(x)\rangle = \pm\,{\bar\sigma}$, with ${\bar\sigma}$ known
exactly \cite{wumccoy},
\be\label{sigmabar}
{\bar\sigma} = {\bar s}\,|m|^{1/8}\,, \qquad  {\bar s} =
2^{1/12}\,e^{-1/8}\,A_{G}^{3/2} = 1.35783834\ldots
\ee
($A_G =1.28242712\ldots$ is Glaisher's constant).
At $\eta = -\infty$ the symmetry is unbroken, $\langle \sigma(x)\rangle =0$.
Another solvable point is $m=0$ (i.e. $\eta=0$), where the theory is not free,
but an exact solution can be obtained due to the presence of infinitely many
local integrals of motion \cite{e8}.

At generic $\eta$ the theory \eqref{ift} is not integrable.
Nonetheless, its particle content is qualitatively understood since \cite{McCoy}.  When
$\eta$ changes, the spectrum of stable particles gradually changes from an infinite
tower of ``mesons'' in the ``low-$T$'' regime at $\eta\to +\infty$ to a single particle
in the``high-$T$'' regime $\eta\to -\infty$. On a quantitative level, various expansions
of some of the masses near the integrable points are available through the perturbation
theory \cite{McCoy,Mussardo1,fz1,fz2,Rut1,delfino,fz3,Rut2}, and rather accurate numerical
data were obtained \cite{Mussardo1, fz1} by means of the ``truncated conformal space approach''
of \cite{alz1}, and by the numerical diagonalization of the lattice transfer-matrix
\cite{Grinza, Caselle}.

Apart from the integrable points, much less is known about the particle scattering.
The numerical method of \cite{alz1} is of limited power in addressing this problem.
While this approach is successful in numerical evaluations of low-energy scattering phases
\cite{alz2}, it is not clear how to apply it to an analysis of high-energy scattering.
At the same time, understanding the scattering amplitudes,
in particular their high-energy behavior, can shed some light on the structure of the
theory. When the parameter $\eta$ changes, some particles lose their stability,
becoming virtual or resonance states. Qualitative changes in the spectrum typically
occur when the masses of some resonance states go to infinity (in the units of the
stable particle masses). Analyzing this phenomenon clearly requires some knowledge
about high-energy scattering.

In this situation, perturbation theory around integrable points is a reasonable
way to begin the analysis. In this work we study the $2\to 2$ scattering amplitude in
the high-T domain $m>0$, using perturbation theory in $h$ to the leading order in $h^2$.
Unfortunately, in its present form, this perturbation theory is not as simple and
transparent as the Feynman diagram technique. It relies on the intermediate-state
decompositions, with the use of well-known exact matrix elements (``form-factors'') of the
field $\sigma(x)$ at $h=0$ \cite{Berg}. One of the technical difficulties of
such ``form-factor perturbation theory'' is that separating disconnected contributions and the ``external leg'' mass corrections is not as straightforward as in the standard Feynman diagrammatic. We bypass these difficulties by using the optical theorem and the associated dispersion relation, expressing the amplitude through the total inelastic cross-section
$\sigma_\text{tot}$. This procedure involves the well-known ambiguity in analytic
terms, which can be fixed once the high-energy asymptotic of the amplitude is known.
To eliminate the ambiguity we calculate this asymptotic directly (again, to the order
$h^2$), using the techniques of \cite{fz2}. It turns out that to this order the high
energy behavior of the $2\to 2$ $S$-matrix element is dominated by the logarithmic term
\be\label{slog}
S_{\,2\to 2} \to -1  + {8h^2\,G_3}\,\big[\log(E/E_{0})-i\pi/4\big] + O(h^4) \quad \text{as}
\quad E\to\infty\,,
\ee
where $E$ is the center of mass energy, and the coefficient $G_3$ is
exactly the third moment of the Euclidean spin-spin correlation function
\be\label{f0}
G_3 = \int_{0}^{\infty}\,r^3 \,\langle \sigma(r) \sigma(0)\rangle \,dr =
(2.3475292186\ldots)\,m^{-15/4}\,.
\ee

The asymptotic behavior \eqref{slog} indicates the logarithmic growth
of the leading term in the $h^2$ expansion of the total inelastic cross
section. The unitarity bound $\sigma_\text{tot}\leqslant 1$ suggests that
\eqref{slog} represents but the first terms of a series in the ``leading
logarithms'' $\big(h^2\,\log(E)\big)^n$. We present argument in support
of this statement, the argument which also shows that the leading logarithms
form the exponential series, resulting in the power-like decay $|S_{2\to 2}| \sim
(E)^{-8G_3\,h^2}$ at large $E$. This behavior suggests that while at low
energies the two-particle scattering is dominated by the elastic $2\to 2$
channel, at large $E$ the scattering goes almost entirely into inelastic
channels, even at small $h^2$.


\section{General properties of $2\to 2$ $S$-matrix element}

In this section we briefly describe the general properties of the
$2\to 2$ scattering matrix element, and the $2\to n$ inelastic
cross-section. For the most part, the content of this section is an
adaptation of the textbook basics of the relativistic $S$-matrix theory
(see e.g. \cite{smatrix}), with minor simplifications specific for 
1+1 kinematics.

For simplicity, and in view of the problem at hand,
we assume that the theory has one neutral particle,
with mass $m$, which we refer to as particle $A$.
The kinematic state of the asymptotic particle is characterized by its
on-shell 2-momentum $p^{\mu}$, conveniently parametrized by the
rapidity $\theta$,
$\ p(\theta) \equiv p^{\mu}(\theta) = \big(m\cosh\theta,\ m\sinh\theta\big)$.
Here we use the notations
\be\label{inout}
| \theta_1, \theta_2, \cdots, \theta_n\rangle_{in (out)}
\ee
for the -$in$ (-$out$) state with $n$ particles $A$, with 
rapidities $\theta_1, \theta_2, \cdots, \theta_n$. We assume the following
normalization of the one-particle states,
\be\label{norm}
\langle\,\theta| \theta'\,\rangle =
(2\pi)\,\delta(\theta-\theta')\,.
\ee

In 2D, the $2\rightarrow 2$ scattering is always purely elastic
(i.e. the momenta of the two outgoing particles are the same as
the momenta of the incoming ones) by total energy-momentum
conservation. Therefore one can write
\begin{align}\label{in2out}
| \theta_1, \theta_2 \,\rangle_{in} &= S_{2\to
2}(\theta_1,\theta_2)| \theta_1, \theta_2\,\rangle_{out} +
\sum_{n=3}^{\infty}\,\int \,\frac{d\beta_{1}}{2\pi}\cdots
\frac{d\beta_{n}}{2\pi}\,\times \nonumber
\\
& \times \frac{(2\pi)^2}{n!}\,
\delta^{(2)}(P_{in}-P_{out}) S_{2\rightarrow n}(\theta_1,
\theta_2|\beta_{1},\cdots, \beta_{n})| \beta_{1}, \cdots,
\beta_{n} \rangle_{out}\,,
\end{align}
where $P_{in}=p(\theta_1)+p(\theta_2)$ and $P_{out}=\sum_{i=1}^n
p(\beta_i)$ are the total 2-momenta of the initial and the final
states, respectively. Equation \eqref{in2out} defines the
$2\rightarrow n$ $S$-matrix elements $S_{2\rightarrow n}$. Our
attention will be mostly on the element $S_{2\rightarrow
2}(\theta_1,\theta_2)$. Relativistic invariance demands that it
actually depends on a single variable, the rapidity difference
$\theta_1 -\,\theta_2$; correspondingly, we will write
\begin{eqnarray}
S_{2\to 2}(\theta_1, \theta_2) = S(\theta_1 -\theta_2)\,.
\end{eqnarray}

\begin{figure}[ht]
\centering
\includegraphics[width=12cm]{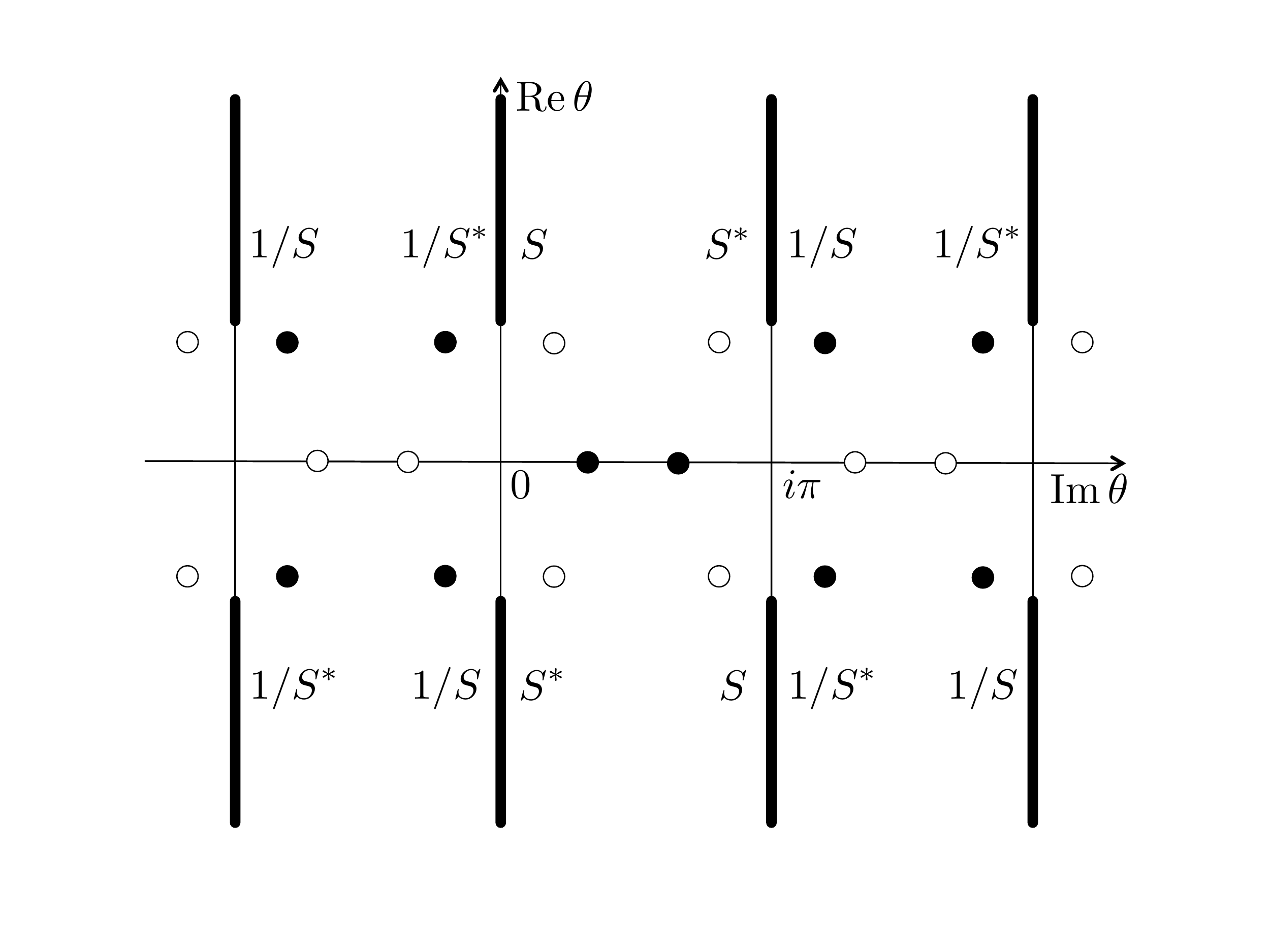}
\caption{Analytic structure of the two-particle scattering
amplitude $S(\theta)$ in the complex $\theta$-plane. The solid
lines represent the branch cuts associated with inelastic channels.
The values of $S(\theta)$ at different edges of the branch cuts
represent physical $S$-matrix element $S$, its complex conjugate $S^*$,
and the inverse values. The bullets $\bullet$ and circles $\circ$
indicate possible positions of poles and zeroes, respectively. Poles
located on the imaginary axis, within the physical strip
$0 < \Im m\, \theta < \pi$ correspond to stable particles;
poles on the strip $-\pi<\Im m\,\theta<0$ are associated with resonance
scattering states.
\label{thetaplane}}
\end{figure}

The function $S(\theta)$ has a direct physical interpretation at real
values of $\theta$ (the physical domain of the $s$-channel scattering),
but it can be analytically continued to the complex $\theta$-plane
with certain singularities.
The analyticity of $S(\theta)$ was widely discussed in the specific
context of factorizable scattering theory (see e.g. \cite{zz1}), but
much of this analysis goes through in the general case. The function
$S(\theta)$ is analytic in the strip
$0 < \Im m \,\theta <\pi$, except for possible poles at the
corresponding segment of the imaginary-$\theta$ axis (such poles, if
present, signify stable particles existing in the theory). $S(\theta)$
takes real values at pure imaginary $\theta$. The
function admits analytic continuation to the whole $\theta$-plane
(with certain branch cuts, as is detailed below) via the functional
relations
\be\label{frel}
S(\theta) = S(i\pi-\theta)\,,
\ee
\be\label{sunit}
S(\theta)S(-\theta)=1\,.
\ee
The first of these relations expresses the crossing symmetry of the
$S$-matrix. The second follows from the unitarity of the $2\to 2$ $S$-matrix
at the energies below the multi-particle thresholds.
In the case of an integrable, purely elastic scattering theory, the multiparticle
contributions
to \eqref{in2out} are altogether absent, and as a result in that
case $S(\theta)$ is a meromorphic function on the whole complex
plane. But in the general (non-integrable) case, of course, branching-point singularities associated with the multiparticle thresholds are present. There are branching points at $\theta =
\pm\,\theta^{(n)}$, where $\theta^{(n)}$ are real positive solutions of
the equation
\be\label{thetan}
\cosh\frac{\theta^{(n)}}{2} = \frac n 2\,;
\ee
they correspond to the $n$-particle thresholds. By crossing
symmetry \eqref{frel}, there are similar branching points
at the axis $\Im m\,\theta = \pi$, at $\theta=i\pi \mp \theta^{(n)}$;
these represent the $n$-particle thresholds in the cross channel.
Since the relations \eqref{frel}, \eqref{sunit} imply periodicity,
\be\label{period}
S(\theta+2i\pi) = S(\theta)\,,
\ee
the pattern is periodically extended along the imaginary-$\theta$
axis. We will define the principal sheet of the $\theta$-surface by
making branch cuts from $\theta^{(n)}$ to $+\infty$ and from $-\theta^{(n)}$
to $-\infty$, with the periodic extension according to Eq.\eqref{period},
as is shown in Fig.~\ref{thetaplane}
(note the unusual directions of the real and imaginary axes in this Figure).
Presumably, $S(\theta)$ can be further analytically
continued to under the branch cuts, to other sheets of the
Riemann surface, but at the moment we do not have much to say about
its analytic structure there. Discontinuities of $S(\theta)$ across the branch
cuts are controlled by the
probabilities of inelastic scattering events. As follows from the full
unitarity condition, we have, at real positive $\theta$
\be\label{fullunit}
S(\theta+i0)S(-\theta+i0) = S(\theta+i0)/S(\theta-i0) =
1 - \sigma_{\rm tot}(\theta)\,,
\ee
where $\sigma_{\rm tot}(\theta)$ is the total probability of all
inelastic processes,
\begin{align}\label{sigmatot}
&\sigma_{\rm tot}(\theta) = \sum_{n=3}^{\infty}\,\sigma_{2\to n}
(\theta)\,, \\
&\sigma_{2\to n}(\theta) = \frac{1}{\sinh\theta}\,\frac{(2\pi)^2}{n!}\,\int\,\frac{d\beta_{1}}{2\pi}\cdots
\frac{d\beta_{n}}{2\pi}\,
\delta^{(2)}(P_{\rm out}-P_{in})\,
\big|S_{2\to n}(\theta_1, \theta_2|\beta_1,\ldots,\beta_{n})\big|^2\,.
\nonumber
\end{align}
In the last line, the notations
are the same as in Eq.\eqref{in2out}\,.
Obviously, $\sigma_{2\to n}(\theta)=0$ at $|\theta| < \theta^{(n)}$, and hence
the total cross-section \eqref{sigmatot} has the segment $[\theta^{(3)},
\infty)$ as its support.

Apart from these branch cuts, $S(\theta)$ can have poles at the principal sheet.
Due to the periodicity \eqref{period}, one can limit attention to the strip
$-\pi \leqslant \Im m \,\theta \leqslant \pi$. We will refer to the domain $0 < \Im m
\,\theta < \pi$ as the ``physical strip'' (PS), since the values of $S(\theta)$
at its boundaries represent physical scattering amplitudes. In view of equation
\eqref{sunit}, every pole of $S(\theta)$ in the PS is accompanied by the associated
zero in the strip $-\pi < \Im m\,\theta < 0$, and vice versa, and for this reason
we refer to the latter as the ``mirror strip'' (MS). By causality, locations of possible poles
within the PS are limited to the imaginary axis $\Re e\,\theta=0$. The poles in the
PS are associated with the stable particles of the theory. In a unitary theory, a pole
at $\theta=i\alpha_p$ (with $0 < \alpha_p <\pi$),
\be\label{poles}
S(\theta) \approx \frac{i r_p}{\theta - i\alpha_p}\,,
\ee
with positive $r_p$ is the direct-channel manifestation of the stable particle
(``bound state'') $A_p$ with the mass $M_p = 2\cos(\alpha_p/2)$. Every
such pole comes along with the cross-channel pole at $i{\tilde\alpha}_p \equiv
i(\pi-\alpha_p)$, with the residue $-ir_p$ \footnote{Of course, strict adherence to our
assumption that there is only one kind of stable particle implies that there can
be only two poles in the PS, one at $i\alpha_A = 2\pi i/3$, and at $i{\tilde\alpha}_A
=i\pi/3$.}. On the other hand, poles in the MS are not restricted to lie on the imaginary
$\theta$-axis. There may be poles at $i\alpha_p \in \text{MS}$ with real $\alpha_p$
(``virtual states'', in the terminology of the potential scattering theory), as well
as with a complex $\alpha_p$ (generally associated with resonance states).

If the positions of all the poles on the principal sheet are known,
the function $S(\theta)$ can be written as
\be\label{blashke}
S(\theta) = \prod_p\,\frac{\sinh\theta + i\sin\alpha_p}{\sinh\theta - i\sin\alpha_p}\,\,U(\theta)\,,
\ee
where the product accounts for all the poles $\theta=i\alpha_p$, $i(\pi-\alpha_p)$, including bound-state, resonance, and virtual
ones. The function $U(\theta)$ satisfies the same relations
\eqref{frel} and \eqref{fullunit} as the amplitude $S(\theta)$,
but in addition it has neither poles nor zeroes on the whole
principal sheet. It can be written as
\be\label{phasedef}
U(\theta) = \exp\big\{i\,\sinh(\theta)\, \Delta(\theta)\big\}\,,
\ee
with $\Delta(\theta)$ analytic everywhere in the PS. As follows
from \eqref{frel} and \eqref{sunit}, $\Delta(\theta)$ satisfies the relations
\be\label{phaseq}
\Delta(\theta) = \Delta(i\pi-\theta)\,, \qquad \Delta(\theta)
=\Delta(-\theta)\,,
\ee
which in turn imply an enhanced periodicity
\be\label{eperiod}
\Delta(\theta+i\pi) = \Delta(\theta)\,.
\ee
Of course, $\Delta(\theta)$ has branching points at the thresholds
$\theta^{(n)}$, and in view of \eqref{fullunit}, its discontinuities
across the associated branch cuts are related to the inelastic
cross-sections. In particular, at positive real $\theta$ we have
\be\label{deltadisct}
\Delta(\theta+i0)-\Delta(\theta-i0) =
-i\,\frac{\log\big(1-\sigma_{\rm tot}(\theta)\big)}{\sinh\theta}\,.
\ee

In view of Eq's \eqref{phaseq}, \eqref{eperiod}, it suffices to focus
on the values of $\Delta(\theta)$ in the strip
$\ 0 \leqslant \Im m\,\theta \leqslant \pi/2$. The variable
transformation
\be\label{wdef}
w = \sinh^2 \theta
\ee
maps this domain (with obvious identifications at the boundary) on the complex
$w$-plane with the branch cut along the real axis,
from $w^{(3)} =\sinh^2\theta^{(3)} = 45/4$ to $+\infty$\footnote{It is useful to
understand the mapping properties of \eqref{wdef}. The edges $\Im m \,\theta = +0$
of the branch cuts $\theta \in [\theta^{(3)},+\infty)$ and $\theta\in(-\infty,-\theta^{(3)}]$ in Fig.~\ref{thetaplane} correspond to the upper and lower edges of
the branch cut in the $w$-plane. The segment $[0,\theta^{(3)}]$ of the imaginary $\theta$-axis is mapped onto the segment $[0,w^{(3)}]$ of the real $w$-axis. Furthermore, 
segments $[-1,0]$ and $(-\infty,-1]$ of the real $w$-axis are the images of segments
$\Re e\,\theta=0\,,\ \Im m\,\theta \in [0,\pi/2]$ and $\Im m\,\theta =\pi/2,\
\Re e\, \theta \in [0,\infty)$, respectively.}. We note that $w$ relates to the center of
mass energy $E=2m\cosh(\theta/2)$ as
\be\label{we}
w = \frac{E^2(E^2-4m^2)}{4m^2}\,.
\ee
With some abuse of notations, let us write
${\Delta}(w)$ and $\sigma_{\rm tot}(w)$ for the quantities defined in Eq's
\eqref{phasedef} and \eqref{sigmatot}, expressed in terms of
the variable $w$. Equation \eqref{deltadisct} then relates the discontinuity
of $\Delta(w)$ across the branch cut in the $w$-plane to $\sigma_\text{tot}(w)$, and
one can write down the associated dispersion relation
\be\label{dispw}
\Delta(w) =  \Delta_\text{reg}(w)+i\,\int_{w^{(3)}}^{\infty}\frac{\log\big(1-\sigma_{\rm tot}(v)\big)}{(w-v)\,\sqrt{v}}\, \frac{dv}{2\pi}\,,
\ee
where $\Delta_\text{reg}(w)$ is an entire function of $w$, real at the real
$w$-axis\footnote{The integral can diverge if $\sigma_\text{tot}$ approaches
its unitarity bound 1 too fast at high energies; as usual, in such cases the appropriate
subtractions are to be made.}.


\section{Scattering amplitude in IFT at weak coupling.}

Let us come back to specific theory \eqref{ift}, and consider particle
scattering in the high-T domain, at weak coupling $|h|\ll m^{15/8}$. In this
domain there is only one stable particle \cite{McCoy}, which at $h=0$ becomes
a free Majorana fermion, in particular $S(\theta)|_{h=0} = -1\ $\footnote{In
1+1 dimensions it is useful to distinguish between $in$- and $out$-states
even in a free fermion theory. Thus, in theory \eqref{ift} with $h=0$,
the $in$-state with $\theta_1 > \theta_2 > \ldots> \theta_n$ is identified  with
the Fock-space state $| \theta_1, \ldots,\theta_n \rangle \equiv a^\dagger
(\theta_1)a^\dagger(\theta_2)\ldots a^\dagger (\theta_n) |0 \rangle$, while for
the corresponding $out$-state the fermion creation operators must be arranged
in the opposite order.}. At small $h$ the amplitude $S(\theta)$ can be
expanded in powers of $h^2$. Here we mostly focus on the leading
term of this expansion, the amplitude $A(\theta)$ in
\be\label{adef}
S(\theta) = -\bigg(1 + {h}^2\,
\frac{iA(\theta)}{\sinh\theta} + O({h}^4)\bigg)\,,
\ee
To simplify notations, in what follows we set the units of mass so that
\be
m=1\,.
\ee

The amplitude $A(\theta)$ is analytic in the PS, except for the poles at
$i\alpha_A=2i\pi/3$ and $i{\tilde\alpha}_A=i\pi/3$, representing particle
$A$ itself in the direct- and the cross-channels, respectively. Note that
in view of the overall minus sign in \eqref{adef}, the residue of $iA(\theta)$ at
$i\alpha_A$ must be positive. As follows from \eqref{frel} and \eqref{sunit},
$\ A(\theta)\, $ satisfies the same symmetries as $\Delta(\theta)$ in
\eqref{phaseq}, i.e.
\be\label{asymm}
A(\theta) = A(-\theta)\,, \quad A(\theta)= A(i\pi-\theta)\,,
\ee
and $A(\theta+i\pi) = A(\theta)$. We will also see that
\be\label{azero}
A(0) =0\,.
\ee
Again, writing $A(w)$ for this amplitude
expressed in terms of the variable \eqref{wdef}, we conclude that this
function is analytic in the whole $w$-plane, with the branch cut from
$w^{(3)}$ to $+\infty$, except for a single pole at $w=-\,3/4$. The discontinuity
of $A(w)$ across the branch cut is related to the cross-section by the well
known optical theorem (a simple consequence of \eqref{fullunit})
\be\label{optical}
A(w+i0)-A(w-i0) = i\sqrt{w}\,\,\sigma_\text{tot}^{(2)}(w)\,,
\ee
where $\sigma_\text{tot}^{(2)}$ is the leading term of the expansion
\be\label{sigmatotexp}
\sigma_\text{tot}(w) = {h}^2\,\sigma_\text{tot}^{(2)}(w)+ O(h^4)\,.
\ee
This leads to the representation
\be\label{adisp}
A(w) = \frac{r\,w}{w+3/4} + w\,\int_{w^{(3)}}^\infty\,\frac{\sigma_\text{tot}^{(2)}(v)
}{(v-w)\,\sqrt{v}}\,\frac{dv}{2\pi}\,,
\ee
where $r$ is a constant, yet to be determined. In writing \eqref{adisp} we have
taken into account the condition \eqref{azero}. We also did not add any analytic
terms, which would be technically consistent with \eqref{optical}. We will
show in Sect.4 below that $A(w) \sim \sqrt{w}\,\log w$ as $w\to\infty$; this asymptotic
(together with \eqref{azero}) rules out such terms.

Of course, the amplitude $A(\theta)$ admits standard perturbation theory representation, as
the integral
\be\label{tproduct}
i\,{A}(\theta_{12}) = -\frac{1}{2}\,\int\,d^2x \,\langle \theta_1,\theta_2|
T \sigma(x)\sigma(0)| \theta_1, \theta_2\rangle_\text{conn}\,,
\ee
where integration is over Minkowski space-time, $x=(\text{x},\text{t})$, and $\theta_{12}$
stands for the rapidity difference $\theta_1-\theta_2$, which we assume to be
non-negative. The integrand
involves the connected part\footnote{The connected part is obtained from the full
matrix element by removing its disconnected parts
\begin{align}\label{4partconnected}
\langle\theta_1,\theta_2|T\sigma(x)\sigma(0)|
\theta_1',\theta_2'\rangle_\text{conn} & = \langle\theta_1,\theta_2|T\sigma(x)\sigma(0)|
\theta_1',\theta_2'\rangle - 2\pi\,\delta(\theta_1-\theta_1')\langle \theta_2|T\sigma(x)\sigma(0)|\theta_2'\rangle- \nonumber\\
&-2\pi\,\delta(\theta_2-\theta_2')\langle \theta_1|T\sigma(x)\sigma(0)|\theta_1'\rangle
- (2\pi)^2\,\delta(\theta_1-\theta_1')\delta(\theta_2-\theta_2')\langle 0 |T\sigma(x)\sigma(0)
|0\rangle
\end{align}
before setting $\theta_1'=\theta_1, \theta_2'=\theta_2$. In writing
\eqref{4partconnected} we have omitted the terms involving
$\delta(\theta_1-\theta_2')$ and $\delta(\theta_1-\theta_2')$ (which are
generally present by the antisymmetry of the matrix element with respect to
$\theta_1 \leftrightarrow \theta_2$) because they vanish at
$\theta_1'=\theta_1$, $\theta_2'=\theta_2$.} of the time-ordered product of the Heisenberg field
operators. In principle, the matrix element in \eqref{tproduct} can be
handled through the intermediate-state decomposition, with the use of the
form-factors
\be\label{2nffactors}
F(\theta_1,\theta_2|\beta_1,\ldots,\beta_{2k+1}) \equiv
\langle \theta_1,\theta_2|\sigma(0)|\beta_1,\ldots,\beta_{2k+1}\rangle\,,
\ee
available in an explicit form \cite{Berg}. Since the form-factors \eqref{2nffactors}
vanish at $\theta_1-\theta_2=0$, this expansion directly confirms the property
\eqref{azero}. However, this approach has well-known difficulties.
In particular, individual $2k+1$-particle contributions are not Lorentz-invariant, this
symmetry being restored only upon summing up all multi-particle terms. For this reason,
it is easier to collect the multi-particle contributions by using the dispersion relation
\eqref{adisp}, expressing the amplitude through the $h^2$ term of the inelastic
cross-section \eqref{sigmatotexp}. The latter can be written in terms of the
form-factors \eqref{2nffactors} as
\begin{align}\label{sigma2tot}
&\sigma_\text{tot}^{(2)}(w) = \sum_{k=1}^{\infty}\,\sigma_{2\to 2k+1}^{(2)}(w)\,,\\
& \sigma_{2\to 2k+1}^{(2)}(w) = \frac{1}{\sqrt{w}}\,
\frac{1}{(2k+1)!}\int \bigg[\prod_{i=1}^{2k+1}\frac{d\beta_i}{2\pi}\bigg]
(2\pi)^2\delta^{(2)}(P_\text{in}-P_\text{out})\,\big|F(\theta_1,\theta_2|\beta_1,\ldots,
\beta_{2k+1})\big|^2\,. \nonumber
\end{align}
This leaves undetermined the coefficient $r$
in the pole term in \eqref{adisp}, which should be evaluated separately.


\subsection{Pole term}

The residue of $A(\theta)$ at the pole $\theta=2i\pi/3$ can be deduced from the
one-particle term in the intermediate-state decomposition of the matrix element
in \eqref{tproduct},
\be\label{onepartterm}
A(\theta_{12}) = -\int_{-\infty}^{\infty}\frac{d\beta\,
\delta(\sinh\theta_1+\sinh\theta_1-\sinh\beta)}
{\cosh\theta_1+\cosh\theta_2-\cosh\beta +i0}\,\big|\langle
\theta_1,\theta_2|\sigma(0)|\beta\rangle\big|^2 + \ldots,
\ee
where $\ldots$ stands for the multi-particle
contributions. With the explicit matrix elements from \cite{Berg},
we have
\be\label{gammasquared}
\big|\langle\theta_1,\theta_2|\sigma(0)|\beta\rangle\big|^2 =
{\bar\sigma}^2\,\tanh^2\bigg(\frac{\theta_1-\theta_2}{2}\bigg)\,
\coth^2\bigg(\frac{\theta_1-\beta}{2}\bigg)\,\coth^2\bigg(\frac{\beta-\theta_2}{2}\bigg).
\ee
Since this expression is positive at all real values of the rapidities, we have
erased the absolute value sign, making possible an analytic continuation to
complex rapidities. The integral over $\beta$ is eliminated by the
delta-function\footnote{It is easiest to do this calculation in the center of mass frame
$\theta_1+\theta_2=0$. Although the one-particle term \eqref{onepartterm} is not Lorentz-invariant, the residue at the pole is.}, and the result explicitly shows a pole at
$\theta\equiv \theta_1-\theta_2=2\pi i/3$,
\be\label{Apole}
A(\theta)\ \approx \ -\, \frac{i\,18\sqrt{3}\,\,{\bar\sigma}^2}{\theta-2\pi i/3}\,.
\ee
Note that the minus sign here appears because analytic continuation
of the r.h.s of \eqref{gammasquared} to the pole point yields the real but
negative value $-\tan^6(2\pi/3) = -27$. This, when combined with the overall minus
sign in \eqref{adef}, Eq.\eqref{Apole} shows $2\pi i/3$ to be a positive pole of
$S(\theta)$,
\be\label{sresidue}
S(\theta)\ \approx \ \frac{i\, 36 ({\bar\sigma} h)^2}{\theta-2\pi i/3} + O(h^4)\,.
\ee
The cross-channel pole at $\theta=i\pi/3$, with the opposite residue, can be extracted from the three-particle contribution to the intermediate-state decomposition, which also generates the threshold singularity at $\theta^{(3)}$. The higher multi-particle terms do not contribute
to the residues, producing only the associated multi-particle threshold singularities.
The residue in \eqref{Apole} corresponds to the value 
\be
r = 36\,{\bar\sigma}^2
\ee
of the coefficient in the pole term in \eqref{adisp}.


\subsection{Three- and multi-particle contributions to $\sigma_\text{tot}^{(2)}$}

\begin{figure}[ht]
\centering
\includegraphics[width=12cm]{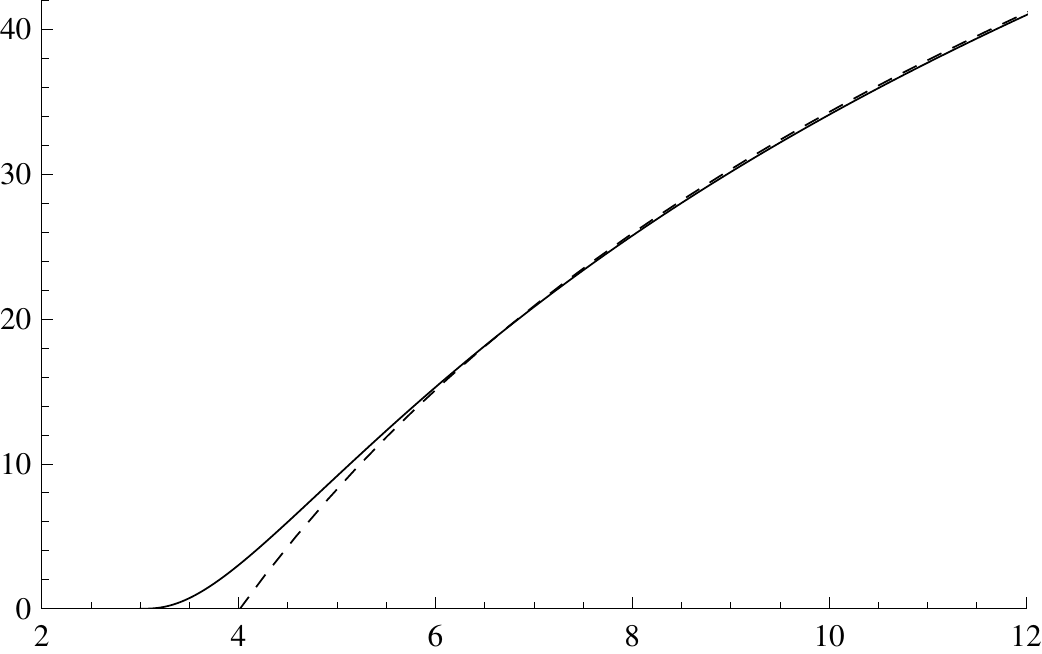}
\caption{Energy dependence of the partial cross-section $\sigma^{(2)}_{2\to 3}(E)$
(solid line). The dashed line shows the asymptotic form \eqref{sigma23ass}, which is seen
to approximate $\sigma^{(2)}_{2\to 3}$ very closely starting from relatively
low energies.\label{PartialCrossSection_LargeE_figure}}
\end{figure}

The first term $\sigma_{2\to 3}^{(2)}$ in \eqref{sigma2tot} can be evaluated in closed
form (see Appendix A for details). When expressed in terms of the center of mass energy
$E = 2\cosh(\theta_{12}/2)$, it
can be written as
\be\label{sigma230}
\sigma_{2\to 3}^{(2)}(E) = \Theta(E-3)\,B(E)\,I(E)\,,
\ee
where the step function $\Theta(E-3)$ is displayed in order to remind that this part of the cross-section vanishes below the three-particle threshold,
\be
B(E)=\frac{4\,{\bar\sigma}^2}{\pi}\,\frac{(E+2)^\frac{5}{2}}{(E-2)^\frac{3}{2}}\,
\frac{(2E-1)^4 (E-3)^3}{(E+1)(E-1)^\frac{5}{2} (E+3)^\frac{3}{2} E^3}\,,
\ee
and $I(E)$ is the elliptic integral
\be\label{Idef}
I(E)= \int_{-1}^{1}\,\bigg(\frac{1-\mu\,t^2}{1-\nu\,t^2}\bigg)^2\,
\frac{\sqrt{1-t^2}}{(1-\lambda \,t^2)^{\frac{3}{2}}}\,dt
\ee
with
\be\label{lmn}
\lambda=\frac{(E+1)(E-3)^3}{(E-1)(E+3)^3}\,, \quad \mu  = \frac{(E-2)(2E+1)^2}{(E+2)(2E-1)^2}\,\lambda\,,
\quad \nu = \frac{E+2}{E-2}\,\lambda\,.
\ee
Some basic properties of this expression are readily derived. Thus, near the
threshold $E=3$
\be
\sigma_{2\to 3}^{(2)}(E) \approx \Theta(E-3)\,\frac{5^6\,\sqrt{5}\,\,{\bar\sigma}^2}{2^5\,3^4\,\sqrt{3}}\,
(E-3)^3\,.
\ee
The integral \eqref{Idef} can be represented by the expansion
\be
I(E) = \frac{\pi}{2}\,\bigg(1 + \frac{2\nu-2\mu+3\lambda}{8} + \ldots\bigg)\,,
\ee
which rapidly converges at any finite $E$. On the other hand, at $E\to\infty$
\be\label{sigma23ass}
\sigma_{2\to 3}^{(2)}(E) = \frac{32\,{\bar\sigma}^2}{\pi}\,\bigg[\log(E^2)+\frac{\sqrt{3}\,\,\pi-11}{2}
+ O\big(1/E^2\big)\bigg]\,.
\ee
The behavior of $\sigma_{2\to 3}^{(2)}(E)$ is shown in Fig.~\ref{PartialCrossSection_LargeE_figure}.

The partial cross-sections $\sigma_{2\to 2k+1}^{(2)}$ with $k>1$ are not evaluated
in closed form. However, numerical estimates show that these multi-particle terms
in \eqref{sigma2tot} are small compared to $\sigma_{2\to 3}^{(2)}$ at all values
of energy $E$. Thus, the five-particle contribution $\sigma_{2\to 5}^{(2)}$
is smaller than 1\% of $\sigma_{2\to 3}^{(2)}$ at all values of $E$, and the
higher terms in \eqref{sigma2tot} are yet much smaller. Therefore, the amplitude $A(\theta)$ in
\eqref{adef} can be approximated with high accuracy by the dispersion relation
\eqref{adisp} with $\sigma_\text{tot}^{(2)}(v)$ replaced by $\sigma_{2\to 3}^{(2)}(v)$.


\subsection{Amplitude}

With $\sigma_\text{tot}^{(2)}(w)$ approximated by $\sigma_{2\to 3}^{(2)}(w)$, the second terms
in \eqref{adisp} (which we denote $A_\sigma (w)$) can be evaluated numerically. The relative
contributions of $A_\sigma(w)$ and the pole term $A_p(w)=36{\bar\sigma}^2\,w/(w+3/4)$
in \eqref{adisp}
\begin{figure}[ht]
\centering
\includegraphics[width=12cm]{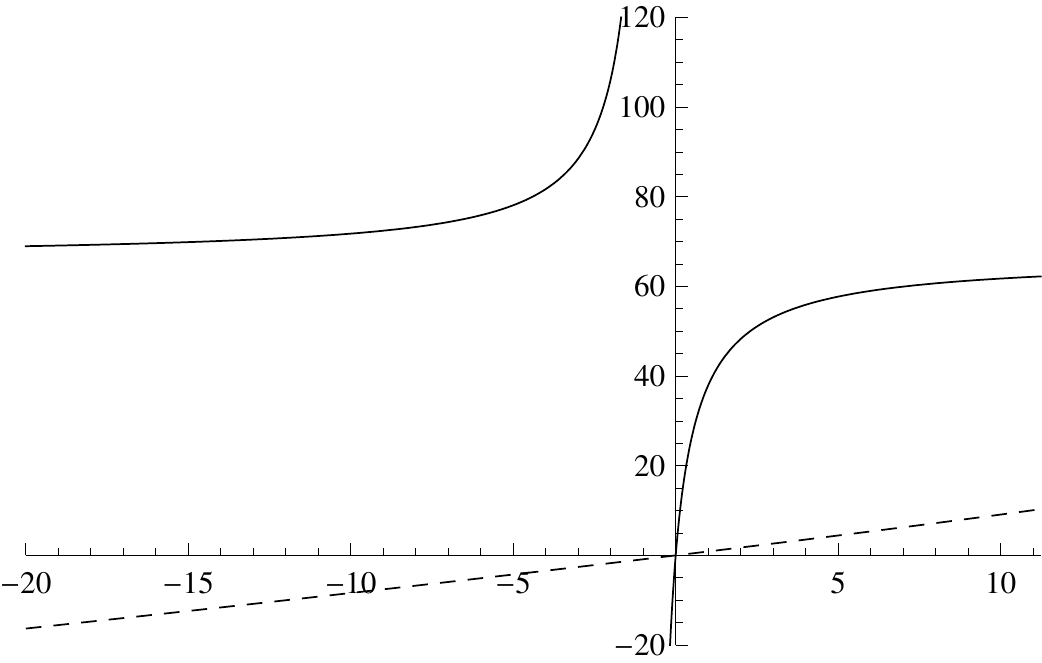}
\caption{Relative contributions of $A_p(w)$ (solid line) and $A_\sigma(w)$ (dashed line) to the amplitude $A(w)$ at low values of $w$.\label{Amplitude_Smallw_figure}}
\end{figure}
are shown in Fig.~\ref{Amplitude_Smallw_figure} for low values of $w$. One can see that in the shown domain of $w$ the
pole term brings a substantially larger contribution then $A_\sigma(w)$. In particular, in the
physical domain $w>0$, but below the three-particle threshold $w^{(3)}=11.25$, $A_\sigma(w)$ contributes
no more then 15\% to the scattering phase. For $w>w^{(3)}$, $A_\sigma(w)$ develops the imaginary
part $\sqrt{w/4}\,\,\sigma_{2\to 3}^{(2)}(w)$. At larger $w$ the real part of $A_\sigma$ increases,
overtaking at $w \approx 68$ the pole term $A_p (w)$ in magnitude, and then approaches
the asymptotic $8{\bar\sigma}^2\,\sqrt{w}$. At large negative $w$, $A_\sigma(w)$ behaves
as $-(8{\bar\sigma}^2/\pi)\,\sqrt{-w}\,\log(-w/w_0)$, with $w_0 =\frac{1}{4}\,\exp\{11-\sqrt{3}\,\,\pi\}
\approx 64.86$.


\section{High energy limit of the amplitude}

In IFT with $h=0$, the matrix element in \eqref{tproduct} admits exact
representation in terms of the Lax equation associated with the special solution
of the classical sinh-Gordon equation. In this section we describe this relation,
and then use it to evaluate the asymptotic of $A(\theta_{12})$ at high energy,
$\theta_{12}\to\infty$.

\subsection{The matrix element}

As is known since the classic paper \cite{McCoy1}, the two-point correlation functions
$G=\langle 0 | T\sigma(x)\sigma(0)| 0\rangle$ and ${\tilde G}=\langle 0 | T \mu(x)
\mu(0)| 0\rangle$ of the spin field $\sigma(x)$ and the disorder field $\mu(x)$ in the $h=0$ IFT
are expressed in terms of the special solution of the classical sinh-Gordon system
\be\label{sinhgordon}
\ptz\ptzb\varphi = \frac{1}{8}\,\sinh(2\varphi)\,,
\quad
\ptz\ptzb\chi = \frac{1}{8}\,\bigg[1-\cosh(2\varphi)\bigg]\,.
\ee
Here and below we use the light-cone coordinates
\be\label{zzbar}
\tz=\tx-\tt\,, \qquad \tzb =\tx+\tt\,
\ee
to label the points $x$ of 1+1 Minkowski space-time. The relevant
solution is Lorentz-invariant, i.e. the functions $\varphi(\tz,\tzb)$,
$\chi(\tz,\tzb)$ depend only on the Lorentz-invariant combination
$\rho=\tz\tzb = \tx^2-\tt^2$. In this case \eqref{sinhgordon} reduces to an ordinary
differential equation, the famous Penlev\'e III equation. The solution
is specified by its $\rho\to 0$ asymptotic form
\be\label{varphichiuv}
\varphi(\rho) = -\frac{1}{2}\,\log\frac{\rho}{4} - \log(-\Omega) + O(\rho^2 \Omega^2)\,,
\quad
\chi(\rho) = \frac{1}{4}\log(16\rho) + \log(-\Omega) + O(\rho)\,,
\ee
where
\be\label{omegadef}
\Omega = \frac{1}{2}\,\log\big(\kappa^2\,\rho\big)\,.
\ee
Under the special choice of the constant $\kappa = e^{\gamma_E}/8$ (which we assume)
$\varphi(\rho)$ is regular at all real $\rho >0$, and decays as
\be\label{varphiir}
\varphi(\rho) = \frac{2}{\pi}\,K_0\big(\sqrt{\rho}\,\big) + O\big(\exp(-3\,\sqrt{\rho}\,)
\big)
\ee
as $\rho\to+\infty$ \cite{McCoy1}. The solution $\varphi(\rho)$ can be regarded as the function of
complex $\rho$. It is possible to show (using e.g. the large-$\rho$ expansions of
Ref.\cite{McCoy1}) that $\varphi(\rho)$ is analytic in the whole $\rho$-plane with the branch
cut from $-\infty$ to $0$, except for the point
$\rho=0$ where the logarithmic singularity (explicit in \eqref{varphichiuv}) is
located. Moreover, the asymptotic form \eqref{varphiir} holds for $-\pi<\text{arg}(\rho)
<\pi$ as $|\rho|\to\infty$. The function $\chi(\rho)$ has similar analyticity, except
that at large $|\rho|$ it tends to a constant,
\be
\chi(\rho) = 4\,\log({\bar s}) + O(e^{-2\sqrt{\rho}})\,, \quad |\rho|\to\infty\,,
\quad -\pi<\text{arg}(\rho)<\pi\,,
\ee
where ${\bar s}$ is the same as in \eqref{sigmabar}.
In the high-$T$ regime ($\langle\sigma\rangle =0\,, \ \langle\mu\rangle = {\bar\sigma}$)
of $h=0$ IFT the correlation functions $G(\rho)$ and ${\tilde G}(\rho)$ are written
in terms of $\varphi(\rho)$ and $\chi(\rho)$ as follows,
\be\label{corrfunctions}
G(\rho) = e^{\chi/2}\,\sinh\big(\varphi/2\big)\,,
\quad {\tilde G}(\rho) = e^{\chi/2}\,\cosh\big(\varphi/2\big)\,.
\ee
The correlation functions are real at space-like separations $\rho>0$, while in the time-like
domain the values at the upper edge of the branch cut, (i.e. $G(\rho+i0)$ and ${\tilde G}(\rho+i0)$)
have to be taken.

This relation can be extended to the matrix elements of the products $T\sigma(x)\sigma(0)$
and $T\mu(x)\mu(0)$ sandwiched between any particle states. The main ingredient is a special
solution $\Psi_{\pm}(x|\beta)$ of the linear system,
\begin{subequations}\label{laxz}
\begin{align}
&\ptz\Psi_{+} = -\frac{1}{2}\,\ptz\varphi\,\Psi_{+}+\frac{1}{4}e^\beta\,e^\varphi
\,\Psi_{-}\,,\label{laxz1}\\
&\ptz\Psi_{-} = \ \ \frac{1}{2}\,\ptz\varphi\,\Psi_{-}-\frac{1}{4}e^\beta\,e^{-\varphi}
\,\Psi_{+}\,,\label{laxz2}
\end{align}
\end{subequations}
and
\begin{subequations}\label{laxzb}
\begin{align}
&\ptzb\Psi_{+} = \ \ \frac{1}{2}\,\ptzb\varphi\,\Psi_{+}-
\frac{1}{4}e^{-\beta}\,e^{-\varphi}\,\Psi_{-}\,,\label{laxzb1}\\
&\ptzb\Psi_{-} = - \frac{1}{2}\,\ptzb\varphi\,\Psi_{-}+
\frac{1}{4}e^{-\beta}\,e^{\varphi}\,\Psi_{+}\,,\label{laxzb2}
\end{align}
\end{subequations}
which constitutes the Lax representation of the sinh-Gordon equation (the first
of Eqs.\eqref{sinhgordon} guarantees integrability of \eqref{laxz1} and \eqref{laxzb1}), with $e^\beta$ playing the role of the spectral
parameter. The solution $\Psi_{\pm}(x|\beta)$ relevant to our problem is described
in some detail in Ref.\cite{fz2} \footnote{In Ref.\cite{fz2} this solution is described
in Euclidean space, where $\tz=\tx+i\ty$, $\tzb=\tx-i\ty$. Here we need its continuation
$\ty \to i\tt$ to Minkowski space-time. The continuation is straightforward for space-like
separations $\rho>0$. In the time-like domain $\rho<0$ the standard $i0$-prescription
$\tz \to \tz + i0\,\text{sign}(\tzb)$, $\tzb\to \tzb+i0\,\text{sign}(\tz)$ is implied.}.
The components $\Psi_\pm(x|\beta)$ are
analytic functions on the double cover of Minkowski space-time parameterized by
the coordinates $x=(\tz,\tzb)$, with branching
singularities at the right and left parts of the light cone, i.e. at $\tz=0$ and $\tzb=0$.
The double cover is needed because $\Psi_{\pm}(x|\beta)$ change sign when the point $x$
is brought around the origin $(\tz,\tzb)=(0,0)$.

Due to the obvious symmetry of Eqs.\eqref{laxz}, \eqref{laxzb}, the functions $\Psi_{\pm}(x|\beta)$ depend only on the combinations $Z=e^\beta\,\tz$, ${\bar Z}=e^{-\beta}\,\tzb$, and below we often use the notation
$\Psi_{\pm}(Z,{\bar Z})$ for them. The solution under consideration can be characterized
by its $(Z,{\bar Z})\to 0$ behavior,
\begin{subequations}
\begin{align}
&\Psi_{+}(Z,{\bar Z}) = \sqrt{\frac{\pi}{-\Omega}}\,\,
\bigg(\frac{Z}{\bar Z}\bigg)^\frac{1}{4}\,\bigg[1+\frac{4\Omega-1}{64}\, {\bar Z}^2 -
\frac{1}{64}\, Z^2 + O(\rho^4 \Omega^2)\bigg]\, ,\\
&\Psi_{-}(Z,{\bar Z}) = \sqrt{\frac{\pi}{-\Omega}}\,\,
\bigg(\frac{\bar Z}{Z}\bigg)^\frac{1}{4}\,\bigg[1+\frac{4\Omega-1}{64}\, Z^2 -
\frac{1}{64}\, {\bar Z}^2 + O(\rho^4 \Omega^2)\bigg]\, ,
\end{align}
\end{subequations}
where $\Omega$ is defined in \eqref{omegadef}.

The matrix elements of the T-product $\sigma(x)\sigma(0)$ between particle states
are certain products of the functions $\Psi_{\pm}(x|\beta)$. The relation looks somewhat simpler
for the ``centered'' product $T \sigma(x/2)\sigma(-x/2)$; this shift of course is irrelevant
in the integral \eqref{tproduct}. First, define the two-particle matrix elements
\begin{subequations}\label{G2}
\begin{align}
&\mathcal{G}(\theta,\theta') = \langle 0 |T \sigma(x/2)\sigma(-x/2)|\theta,\theta'\rangle\,,\\
&{\mathcal G}(\theta|\theta') =\langle \theta |T \sigma(x/2)\sigma(-x/2) | \theta'\rangle
- 2\pi\,\delta(\theta-\theta')\,G\,,
\end{align}
\end{subequations}
where $G=G(\rho)$ is the two-point function \eqref{corrfunctions}, and we omit the argument $x$ in
$\mathcal{G}(\theta,\theta')$ and $\mathcal{G}(\theta|\theta')$ to simplify notations.
Then we have
\begin{subequations}\label{G12}
\begin{align}
\mathcal{G}(\theta_1,\theta_2) &=-\frac{i}{2}\,
\bigg[G\,\,\frac{e^{\theta_1}-e^{\theta_2}}{e^{\theta_1}+e^{\theta_2}}
\,\Psi_{s}(\theta_1,\theta_2) - {\tilde G}\,\Psi_a (\theta_1,\theta_2)\bigg]\,,\\
\mathcal{G}(\theta_1|\theta_2) &= -\frac{1}{2}\,
\bigg[G\,\,\frac{e^{\theta_1}+e^{\theta_2}}{e^{\theta_1}-e^{\theta_2}}
\,\Psi_{a}(\theta_1,\theta_2) - {\tilde G}\,\Psi_s (\theta_1,\theta_2)\bigg]\,,
\end{align}
\end{subequations}
where $G$ and ${\tilde G}$ are the two-point functions \eqref{corrfunctions}, and
\begin{subequations}
\begin{align}
&\Psi_s (\theta_1,\theta_2) = \Psi_{+}(Z_1,{\bar Z}_1)\Psi_{-}(Z_2,{\bar Z}_2)+
\Psi_{-}(Z_1,{\bar Z}_1)\Psi_{+}(Z_2,{\bar Z}_2)\,,\\
&\Psi_a (\theta_1,\theta_2) = \Psi_{+}(Z_1,{\bar Z}_1)\Psi_{-}(Z_2,{\bar Z}_2)-
\Psi_{-}(Z_1,{\bar Z}_1)
\Psi_{+}(Z_2,{\bar Z}_2)\,.
\end{align}
\end{subequations}
Here
\be
(Z_1,{\bar Z}_1) = \big(e^{\theta_1}\tz,\, e^{-\theta_1}\tzb\big)\,,
\quad (Z_2,{\bar Z}_2) = \big(e^{\theta_2}\tz,\, e^{-\theta_2}\tzb\big)\,.
\ee
Finally, the four-particle matrix element in \eqref{tproduct} is the fermionic Wick product of the two-particle matrix elements \eqref{G12},
\begin{align}\label{conndef}
&\langle \theta_1,\theta_2 |T \sigma(x/2)\sigma(-x/2) |
\theta_1, \theta_2\rangle_\text{conn} = \nonumber
\\
&=G^{-1}\big[\mathcal{G}(\theta_2|\theta_1)\mathcal{G}(\theta_1|\theta_2)
-\mathcal{G}(\theta_1|\theta_1)\mathcal{G}(\theta_2|\theta_2) +
\mathcal{G}(\theta_1,\theta_2)\mathcal{G}(\theta_1,\theta_2)\big].
\end{align}

It will be convenient to write $\mathcal{G}(\theta|\theta)$ (which is just the limit
$\theta'=\theta$ of $\mathcal{G}(\theta|\theta')$) as the sum of two terms
\be\label{Gdiag}
\mathcal{G}(\theta|\theta) = {\tilde G}\,\,K(Z,{\bar Z})-G\,\,L(Z,{\bar Z}) \,,
\ee
where
\begin{subequations}\label{KLdef}
\begin{align}
& K(Z,{\bar Z}) = \Psi_{+}(Z,{\bar Z})\Psi_{-}(Z,{\bar Z})\,,
\\
& L(Z,{\bar Z}) = \Psi_{-}(Z,{\bar Z})\partial_\theta\Psi_{+}(Z,{\bar Z})-
\Psi_{+}(Z,{\bar Z})\partial_\theta\Psi_{-}(Z,{\bar Z})\,,
\end{align}
\end{subequations}
and again $Z=e^\theta \tz$, ${\bar Z}=e^{-\theta} \tzb$.


\subsection{High energy limit of the integral \eqref{tproduct}}

The expression for the matrix element in \eqref{tproduct} described above is
still too complicated for the integral to be evaluated in closed form. However,
it is possible to use this representation to derive its limiting behavior at
$\theta_{12}\to\infty$.

To simplify arguments, we assume that $\theta_1 \to+\infty$
and $\theta_2\to-\infty$. In this limit the factors involving the ratio of $e^{\theta_1}-
e^{\theta_2}$ and $e^{\theta_1}+e^{\theta_2}$ in \eqref{G12} can be dropped, and
the expression \eqref{conndef} simplifies as
\begin{align}\label{G4uv}
\langle \theta_1,\theta_2 |T \sigma(x/2)\sigma(-x/2) |
\theta_1, \theta_2\rangle_\text{conn}  \to &\,
G\,[L(Z_1,{\bar Z_1})L(Z_2,{\bar Z}_2)+K(Z_1,{\bar Z_1})K(Z_2,{\bar Z}_2)]- \nonumber
\\
&-{\tilde G}\,[K(Z_1,{\bar Z_1})L(Z_2,{\bar Z}_2)+L(Z_1,{\bar Z_1})K(Z_2,{\bar Z}_2)],
\end{align}
where $K$ and $L$ are the functions \eqref{KLdef}. Furthermore, it is possible to
show that at $|Z+{\bar Z}| \gg 1$ the functions $\Psi_{\pm}(Z,{\bar Z})$ approach
plane waves. The exact form looks different for different regions in the $(Z,{\bar Z})$-plane.
Thus, at $Z\to +\infty$, and $Z \gg {\bar Z}$ we have
\be\label{psiZp}
\Psi_{+}(Z,{\bar Z}) \to 2\,e^{\varphi/2}\,\cos\bigg(\frac{Z-{\bar Z}}{4}-\frac{\pi}{4}\bigg)\,,
\quad \Psi_{-}(Z,{\bar Z}) \to 2\,e^{-\varphi/2}\,\cos\bigg(\frac{Z-{\bar Z}}{4}+\frac{\pi}{4}\bigg)\,,
\ee
while at ${\bar Z}\to +\infty$ and ${\bar Z}\gg Z$
\be\label{psiZbp}
\Psi_{+}(Z,{\bar Z}) \to 2\,e^{-\varphi/2}\,\cos\bigg(\frac{Z-{\bar Z}}{4}-\frac{\pi}{4}\bigg)\,,
\quad \Psi_{-}(Z,{\bar Z}) \to 2\,e^{\varphi/2}\,\cos\bigg(\frac{Z-{\bar Z}}{4}+\frac{\pi}{4}\bigg)
\ee
(when $Z\sim{\bar Z} \to +\infty$ the difference between \eqref{psiZp} and \eqref{psiZbp}
is irrelevant since in this domain $\varphi\to 0$). Likewise, at $Z\to -\infty$ and
$-Z \gg{\bar Z}$
\be\label{psiZm}
\Psi_{+}(Z,{\bar Z}) \to 2i\,e^{\varphi/2}\,\cos\bigg(\frac{Z-{\bar Z}}{4}+\frac{\pi}{4}\bigg)\,,
\quad \Psi_{-}(Z,{\bar Z}) \to -2i\,e^{-\varphi/2}\,\cos\bigg(\frac{Z-{\bar Z}}{4}-\frac{\pi}{4}\bigg),
\ee
while at ${\bar Z}\to -\infty$ and $-{\bar Z}\gg Z$
\be\label{psiZbm}
\Psi_{+}(Z,{\bar Z}) \to 2i\,e^{-\varphi/2}\,\cos\bigg(\frac{Z-{\bar Z}}{4}+\frac{\pi}{4}\bigg)\,,
\quad \Psi_{-}(Z,{\bar Z}) \to -2i\,e^{\varphi/2}\,\cos\bigg(\frac{Z-{\bar Z}}{4}-\frac{\pi}{4}\bigg)\,.
\ee
In writing these asymptotics we have ignored the overall sign ambiguity of $\Psi_{\pm}$ (if the sign at $Z+{\bar Z} \to +\infty$ is fixed as in Eqs.\eqref{psiZp}, \eqref{psiZbp}, the overall signs in \eqref{psiZm}, \eqref{psiZbm} still depend on the way of continuation around the point $(Z,{\bar Z})=(0,0)$), since this ambiguity does not affect functions \eqref{KLdef}. From these equations one finds the corresponding asymptotics of functions \eqref{KLdef},
\be
K(Z,{\bar Z}) \to 2\,\cos\bigg(\frac{Z-{\bar Z}}{2}\bigg)\,, \quad
L(Z,{\bar Z}) \to |Z+{\bar Z}|
\ee
as $|Z+{\bar Z}| \to\infty$.

The leading $\theta_{12}\to\infty$ asymptotic of the integral \eqref{tproduct}
clearly comes from the first term, $G\,L(Z_1,{\bar Z}_1)L(Z_2,{\bar Z}_2)$, in \eqref{G4uv}.
The remaining terms involve the fast oscillating function $K$, and integration over $(\tz,\tzb)$
leads to their suppression by at least one power of $e^{-\theta_{12}}$, as compared
to the leading term. Thus, the leading high energy asymptotic $\theta_{12}\to\infty$ of the
amplitude \eqref{tproduct} is given by
\be\label{aassint}
{iA(\theta_{12})} \to -\,\frac{1}{2}\,\int\,d^2 x \,|UV|\,G(\rho+i0), \quad
\theta_{12}\to +\infty\,,
\ee
where $d^2 x = d\tx d\tt = \frac{1}{2}\,d\tz d\tzb$, and
\be\label{UVdef}
U = Z_1 +{\bar Z}_1 = e^{\theta_1}\,\tz + e^{-\theta_{1}}\,\tzb\,,\qquad
V = Z_2 +{\bar Z}_2 = e^{\theta_2}\,\tz + e^{-\theta_{2}}\,\tzb\,.
\ee
The integral in \eqref{aassint} can be handled as follows. Write
\be\label{modsplit}
|UV| = UV - 2UV\,\Theta(-UV)\,,
\ee
where $\Theta(x)$ is the usual step function. The first term here is analytic in the
coordinates $(\tz,\tzb)$, and its contribution can be evaluated using the Wick rotation
$\tt\to-i\ty$, yielding
\be
-\,\frac{1}{2}\,\int\,d^2 x\,UV\,G(\tz\tzb) \to  \pi\,i\,G_{3}\,\exp(\theta_{12})\,
\quad \text{as}\quad \theta_{12}\to +\infty\,,
\ee
where
\be
G_{3} = \frac{1}{2}\,\int_{0}^{\infty}\,\rho\,G(\rho)\,d\rho
\ee
is the third moment of the Euclidean spin-spin correlation function \eqref{f0}. The
second term in \eqref{modsplit} involves the step function which limits the integration
domain to $UV <0$. In the domain $U<0,\ V>0$ (which lies entirely within the future light
cone $\rho = \tz\tzb < 0\,,\ \tz<0$) one can parameterize $\tz = -\sqrt{-\rho}\,\,
e^{-\phi}\,, \tzb = \sqrt{-\rho}\,\,e^{\phi}$, with the integration domain corresponding
to the range $\theta_2 < \phi < \theta_1$ of the Lorentz parameter $\phi$.
Therefore, at $\theta_{12}\to +\infty$ the contribution from this domain to \eqref{aassint} is
\begin{align}\label{UVintr}
\int_{U<0<V}\,UV\,G(\rho+i0)\,d^2 x \to -\,\theta_{12}\,&\exp(\theta_{12})\,
\int_{-\infty}^{0}\,\rho\,G(\rho+i0)\,d\rho= \nonumber\\
&=\theta_{12}\,\exp(\theta_{12})\,
\int_{0}^{\infty}\,\rho\,G(\rho)\,d\rho\,,
\end{align}
where the last form is obtained by rotating the the integration contour $\rho\to e^{-i\pi}\rho$,
which is possible since $G(\rho)$ decays exponentially into the upper half-plane of complex $\rho$. The ``past'' part of the light cone $\tz >0>\tzb$ brings in an identical contribution. Collecting
all these pieces together, we have
\be\label{aass}
\frac{iA(\theta_{12})}{\sinh\theta_{12}} \to -4\,G_{3}\,\,\big[\theta_{12}-i\pi/2 + \theta_0\big]\,,
\ee
where $\theta_0$ is a real constant whose value can not be determined by the above simple analysis
(its calculation would require much better understanding of the behavior of $L(Z,{\bar Z})$ in the
domain $Z +{\bar Z} \sim 1$).

The asymptotic \eqref{aass} is written in terms of the variable \eqref{wdef} as
\be\label{awass}
iA(w+i0) \to -2\,G_3\,\sqrt{w}\,\big[\log(w/w_0)-i\pi\big]\,, \quad w \to +\infty\,,
\ee
where $w_0$ is a real positive constant related to $\theta_0$ in \eqref{aass}. This equation
can be extended to the whole complex $w$-plane with the branch cut along the positive
part of the real axis,
\be
A(w) \to - 2\,G_3\,\sqrt{-w}\,\,\log(-w/w_0), \quad |w|\to\infty\,, \quad
-\pi<\text{arg}(-w)<\pi\,,
\ee
where the branch of the multi-valued function is chosen in such a way that
$\sqrt{-w}$ is positive and $\log(-w)$ is real at real negative $w$. From
\eqref{optical} one finds the high-energy behavior of the $\sim h^2$ term
in the inelastic cross-section
\be\label{sigma2log}
\sigma_\text{tot}^{(2)} (w) \to 4\,G_3 \,\log(w/w_0)  \sim 8\,G_3\,
\log(E^2)\,,
\ee
which is of the same logarithmic form as the high energy behavior of
$\sigma^{(2)}_{2\to 3}$, determined in sect.3, Eq.\eqref{sigma23ass}. Note
that the coefficient in \eqref{sigma23ass} corresponds to the 1-particle
contribution $G_\text{1-part}(\rho) = (1/\pi)\,K_0 (\sqrt{\rho})$ to the
intermediate-state decomposition of the spin-spin correlation function $G(\rho)$,
\be
\frac{1}{2}\,\int_{0}^\infty \,
\rho\,G_\text{1-part}(\rho)\,d\rho\ = \ \frac{4{\bar\sigma}^2}{\pi}\ =\ 2.3475035314\ldots
\ee
The smallness of the difference between the exact value of $G_3$ in \eqref{f0} and
the one-particle contribution reaffirms the statement that $\sigma_\text{tot}^{(2)}$
is dominated by its three-particle component $\sigma_{2\to 3}^{(2)}$.


\section{Remark on higher orders in $h^2$}

The logarithmic growth \eqref{sigma2log} of the $h^2$ term in the inelastic
cross-section,
\be
\sigma_\text{tot} \to 8\,G_3\,h^2\,\log(E^2) +O(h^4) \quad \text{as}\quad E\to \infty\,,
\ee
suggests that the higher-order terms of the $h^2$ expansion become
significant at high energies, since unitarity requires $\sigma_\text{tot}
\leqslant 1$ at all energies. One can expect that the terms of the order $h^{2n}$
in the expansion \eqref{sigmatotexp} are $\sim h^{2n}\,\log^n(E^2)$ at large
$E$. Here we argue that this is indeed the case, and moreover, that summing up
these leading logarithms results in a power-like behavior\footnote{This
behavior, with $\alpha = 64\,({\bar\sigma}h)^2/(\pi\,m^4)$, was earlier
proposed by S.~Rutkevich \cite{rutprivate} on the basis of a different argument.
He also conjectured the high-energy form of the partial cross-sections
$\sigma_{2\to {2+n}} \approx (1/n!)\,(\alpha\,\log(E/E_0))^n\,(E/E_0)^{-\alpha}$.}
\be\label{sigmaEass}
\sigma_\text{tot} \to 1 - (E/E_0)^{-\alpha}
\ee
at $E\gg1$, where the exponent $\alpha \approx 16\,G_3\,h^2$  at small $h^2$,
and $E_0 = \text{Const}\,m$, with yet unknown constant.

First, let us note that the expression \eqref{aassint} admits a simple semiclassical 
interpretation, as follows. Let us introduce coordinates
\be\label{uvdef}
(u,v) = (e^{\theta_1}\,\tz + e^{-\theta_1}\,\tzb, \,\,
e^{\theta_2}\,\tz + e^{-\theta_2}\,\tzb)
\ee
on the $(\tz,\tzb)$-plane, and assume that at $\theta_{12}\to\infty$ the
particles with the rapidities $\theta_1$ and $\theta_2$ can be represented
by their classical trajectories $u=0$ and $v=0$, respectively. Also assume
that unless the points $x_1$, $x_2$ are too close to the trajectories, the matrix element
$\langle \theta_1,\theta_2| T\sigma(x_1)\sigma(x_2)| \theta_1,\theta_2\rangle$ is
approximated by the correlation function $\langle 0| T\sigma(x_1)\sigma(x_2)| 0\rangle$,
multiplied by the sign factor which depends on whether the insertion points $x_1$ and $x_2$
are located on the same side or on different sides of each of the trajectories. More
specifically, we assume\footnote{To simplify notations, here and below we ignore the delta functions
associated with the plane-wave normalization \eqref{norm} of the particle states. Instead,
one can think of normalized wave packets centered at the rapidities $\theta_1$ and $\theta_2$.}
\begin{eqnarray}
\langle \theta_1,\theta_2| T\sigma(x_1)\sigma(x_2)| \theta_1,\theta_2\rangle
\approx \text{sign}(u_1)\, \text{sign}(u_2)\,\text{sign}(v_1)\,\text{sign}(v_2)\,\langle 0| T\sigma(x_1)\sigma(x_2)| 0\rangle\,,
\end{eqnarray}
where $(u_1,v_1)$ and $(u_2,v_2)$ are the coordinates \eqref{uvdef} of the
points $x_1$ and $x_2$, respectively\footnote{A similar approximation was used in Ref.\cite{sachdev}, in a different context.
In our case it can be justified
as follows. The correlation between the spin operators located at the points $x_1$ and
$x_2$ is due to exchanges by odd numbers of particles. An external particle trajectory
which passes between these two points intersects an odd number of trajectories of these
exchange particles, thus producing the minus sign.}.
This expression still involves the ``disconnected parts'' $[1+(\text{sign}(u_1)\,\text{sign}(v_1)-1)+(\text{sign}(u_2)\,\text{sign}(v_2)-1)]\,
\langle 0| T\sigma(x_1)\sigma(x_2)| 0\rangle$. Subtracting those, one obtains
\be
\langle \theta_1,\theta_2| T\sigma(x_1)\sigma(x_2)| \theta_1,\theta_2\rangle_\text
{conn}
\approx 4\,\Theta(-u_1 v_1)\Theta(-u_2 v_2)\,\langle 0| T\sigma(x_1)\sigma(x_2)| 0\rangle\,,
\ee
where again $\Theta(u)$ is the conventional step function. Since the correlation
function $\langle 0| T\sigma(x_1)\sigma(x_2)| 0\rangle$ depends only on the
separation $x_1-x_2$, one can explicitly integrate out one of the coordinates,
\begin{align}\label{aint1}
-\frac{1}{2}\,\int\,d^2 x_1\,d^2 x_2 \,&
\langle \theta_1,\theta_2| T\sigma(x_1)\sigma(x_2)| \theta_1,\theta_2\rangle_\text
{conn} =\nonumber
\\
&= -\frac{1}{2\,\sinh\theta_{12}}\,
\int\,d^2 x \,|UV|\,\langle 0| T\sigma(x/2)\sigma(-x/2)| 0\rangle\,,
\end{align}
where $U=u_1-u_2$, $V=v_1-v_2$ are the coordinates \eqref{uvdef} associated with the
separation $x=x_1-x_2$. The last expression is equivalent to \eqref{aassint}, which leads
to logarithmic behavior \eqref{awass}. The origin of the logarithm is clear. Apart from the factor $|UV|$, the integrand in \eqref{aint1} is Lorentz-invariant. The step function in \eqref{modsplit} provides a cutoff for the integration over the Lorentz boost parameter $\phi$ in the integral over
$d^2 x = d\rho d\phi$. The factor $\theta_{12} \approx \frac{1}{2}\,\log w$ in
\eqref{UVintr} is just the size of the Lorentz boost which brings the rest frame
of particle 2 to the rest frame of particle 1.

The same approximation can be applied to the higher-order terms in the usual
perturbation theory expansion of
\be\label{stproduct}
S(\theta_{12}) = - \,\langle\theta_1,\theta_2| T\exp\big\{-ih\,\int\,\sigma(x)\,d^2 x\big\}
| \theta_1,\theta_2\rangle_\text{conn}
\ee
by assuming that at $\theta_{12}\to \infty$ one can write
\begin{align}\label{2npoint}
\langle\theta_1,\theta_2| T\sigma(x_1)\sigma(x_2)\ldots&\sigma(x_{2n})|\theta_1,
\theta_2\rangle \to \nonumber 
\\
\to\bigg[\prod_{i=1}^{2n} &\,\text{sign}(u_i)\,\text{sign}(v_i)\bigg]\,
\langle 0| T\sigma(x_1)\sigma(x_2)\ldots\sigma(x_{2n})| 0\rangle\,,
\end{align}
where $(u_i,v_i)$ are the coordinates \eqref{uvdef} of the points $x_i$. Again,
apart from the sign factors $\text{sign}(u_i)\,\text{sign}(v_i)$, the r.h.s.\ of \eqref{2npoint} is
invariant with respect to simultaneous Lorentz transformations of the coordinates
$x_i$. Therefore, integration over $\prod_{i=1}^{2n}\,d^2 x_i$ generates at least
one factor $\theta_{12}$. But it is easy to see that this integration in fact produces a
much larger contribution $\sim \theta_{12}^n$. Indeed, the $2n$-point correlation function
contains $(2n)!/(2^n\,n!)$ fully disconnected terms
\be
\langle 0| T\sigma(x_1)\sigma(x_2)| 0\rangle \langle 0| T\sigma(x_3)\sigma(x_4)| 0\rangle
\ldots\langle 0| T\sigma(x_{2n-1})\sigma(x_{2n})| 0\rangle + \text{permutations}\,.
\ee
These terms are invariant with respect to $n$ copies of the Lorentz group, each
acting on a respective pair of the points. Because of the factorization, these terms
can be handled exactly as it was done above with the $\sim h^2$ contribution, and
lead to the contributions $\sim (h^2\,\theta_{12})^n$ to
\eqref{stproduct}. It is not difficult to see that the contributions from the above
fully disconnected terms form exponential series in $h^2\,\theta_{12} \sim h^2\,\log w$.
Assuming that there are no other sources of these ``leading logarithms'' (which of course
needs to be proven by more detailed analysis), this suggests the power-like
decay of the $2\to 2$ $S$-matrix element in terms of the variable \eqref{wdef},
\be\label{sassw}
S(w) \sim (-w)^{-2G_{3} h^2} \quad \text{as}\quad |w|\to \infty \quad
(-\pi <\text{arg}(w) < \pi)\,,
\ee
where again the branch of the function is such that $S(w)$ is real at negative
$w$. This corresponds to the behavior \eqref{sigmaEass} of the inelastic cross-section.

Finally, we note that the asymptotic behavior \eqref{sassw} agrees with naive
``two-particle unitarization'' of \eqref{adef},
\be\label{2punit}
S(\theta) = - \,\frac{\sinh\theta + i\sin(2\pi/3)}{\sinh\theta - i\sin(2\pi/3)}
\,\frac{\sinh\theta - i\sin(2\pi/3+rh^2)}{\sinh\theta + i\sin(2\pi/3+r h^2)}\,
\exp\bigg(\frac{ih^2\,{ A}_\sigma(\theta)}{\sinh\theta}\bigg)\,,
\ee
where $r=36{\bar\sigma}^2$, and ${A}_\sigma(\theta)$ is the second term in
\eqref{adisp}\footnote{Although \eqref{2punit} involves all powers of $h^2$, it of
course is not an exact expression for the $2\to 2$ $S$-matrix element, since it
ignores the higher-order corrections to the residue \eqref{sresidue}, as well
as the higher order terms in $\sigma_\text{tot}$.}.


\section*{Appendix}

\appendix

\section{Calculation of $\sigma_{2\to 3}^{(2)}$}

To simplify notations, we will perform the calculations in the center of mass frame
$\theta_1=-\theta_2=\theta/2$, and assume that $\theta >0$. It is useful to introduce
notation $x = e^{\theta/2}$ and $E=x+x^{-1}$; the last one has the meaning of
the center of mass energy. Kinematics demands that $E\geqslant 3$, i.e.
$x \geqslant (3+\sqrt{5})/2$. We have
\be
\sqrt{w} = \frac{x^4-1}{2x^2}.
\ee
The relevant $2\to 3$ matrix elements \eqref{2nffactors} have explicit form~\cite{Berg}
\be\label{ffact23}
F(\theta_1,\theta_2|\beta_1,\beta_2,\beta_3) = \frac{x^2-1}{x^2+1}\,G(x|\{z_i\})\,,
\ee
with
\be
G(x|\{z_i\}) = {\bar\sigma}\,\prod_{k=1}^3 \frac{x+z_k}{x-z_k}\frac{1+x z_k}{1-x z_k}\,\prod_{i<j}^3
\frac{z_i-z_j}{z_i+z_j}\,,
\ee
where $z_i =e^{\beta_i}$, $i=1$, 2, 3. The expression \eqref{sigma2tot} with $k=1$ takes
the form
\be\label{sigmaW}
\sigma_{2\to 3}^{(2)} = \frac{2\,{\bar\sigma}^2}{\pi}\,\frac{x^2(x^2-1)}{(x^2+1)^3}\,\,W(x) =
\frac{2\,{\bar\sigma}^2}{\pi}\,\frac{\sqrt{E^2-4}}{E^3}\,\,W(x)\,,
\ee
where
\be\label{Wint}
W(x)=\frac{1}{3!}\,\int\,\frac{dz_1}{z_1}\frac{dz_2}{z_2}\frac{dz_3}{z_3}\,
\delta(E-z_1-z_2-z_3)\delta(E-1/z_1-1/z_2-1/z_3)\,G^2(x|\{z_i\})\,.
\ee
The integration in \eqref{Wint} is from $0$ to $\infty$ for each of the variables $z_i$.
Since the integrand is obviously symmetric with respect to permutations of these variables,
one can eliminate the factor $1/3!$ by limiting the integration to the domain
$0<z_1<z_2<z_3<\infty$. The integral can be further simplified by changing to
symmetric variables
\be\label{sdef}
s_1=z_1+z_2+z_3\,,\quad s_2=z_1 z_2+z_1 z_3+ z_2 z_3\,, \quad s=z_1 z_2 z_3
\ee
with the Jacobian
\be\label{Ddef}
D = \bigg|\frac{\partial(s_1,s_2,s)}{\partial(z_1,z_2,z_3)}\bigg| =
(z_2-z_1)(z_3-z_2)(z_3-z_1)
\ee
having the following form in terms of the variables \eqref{sdef},
\be\label{Dform1}
D = \sqrt{s_1^2 s_2^2 -4\,s_1^3 s - 4\,s_2^3 + 18\, s_1 s_2 s - 27\,s^2}\,\,.
\ee
The integrations over $s_1$ and $s_2$ are eliminated by the delta-functions
in \eqref{Wint}, which set
\be\label{sconstraint}
s_1=E\,, \qquad  s_2=s\,E\,.
\ee
At these values the Jacobian simplifies as
\be
D \to D(E|s)= 2E^\frac{3}{2}\,\sqrt{s(s-s_{-})(s_{+}-s)}\,\,,
\ee
where $s_{\pm}$ are ordered ($s_{+}\geqslant s_{-}$) roots of the quadratic polynomial
\be
-P(E|s) = s^2 - \frac{E^4+18 E^2 -27}{4 E^3}\,s +1\,.
\ee
Note that at $E\geqslant 3$ the roots are real and positive, and that they collide
at the three-particle threshold $E=3$. The remaining integral over $s$ takes the
form
\be\label{Ws}
W = \int_{s_{-}}^{s_{+}}\,\frac{G^2(E|s)}{D(E|s)}\,d s\,,
\ee
where $G(E|s)$ is the factor $G(x|\{z_i\})$ in \eqref{ffact23} expressed in terms
of the variables \eqref{sdef}, and specified to the values \eqref{sconstraint}.
Explicit calculation yields
\be
G(E|s) = \frac{2E^2+1}{E^2-1}\,\frac{R(E|s)}{Q(E|s)}\,\frac{D(E|s)}{s}\,,
\ee
where
\be
R(E|s)=s^2+\frac{E(4E^2-7)}{2E^2+1}\,s + 1\,, \qquad Q(E|s)=s^2-E\,s+1\,.
\ee

To simplify analysis of the integral \eqref{Ws}, it is useful to make a projective
transformation of the integration variable to bring the limits to $-1$ and $+1$.
We define the parameter
\be
\lambda = \lambda(E) = \bigg(\frac{s_{+}-1}{s_{+}+1}\bigg)^2=\frac{(E+1)(E-3)^3}{(E-1)(E+3)^3}\,\,,
\ee
and trade $s$ for the variable
\be
t = \frac{1}{\sqrt{\lambda}}\,\frac{1-s}{1+s}\,.
\ee
This brings \eqref{Ws} to the form
\be\label{Wform}
W = \frac{2\,(E+2)^2(2E-1)^4(E-3)^3}{(E-2)^2(E^2-1)(E+3)^{\frac{3}{2}}(E-1)^{\frac{3}{2}}}\,
\int_{-1}^{1}\,\bigg(\frac{1-\mu\,t^2}{1-\nu\,t^2}\bigg)^2\,
\frac{\sqrt{1-t^2}}{(1-\lambda t^2)^{\frac{3}{2}}}\,dt\,,
\ee
where $\mu=\mu(E)$ and $\nu = \nu(E)$ are given in Eq.\eqref{lmn}. Note that
$0< \mu < \lambda <\nu < 1$ at all $E>3$, and at $E\to\infty$ these parameters
approach $1$. Combining \eqref{Wform} and \eqref{sigmaW}, one arrives at \eqref{sigma230}.


\bigskip
\section*{Acknowledgments}

AZ is grateful to S.~Rutkevich for many important communications on the subject.
We would like to thank S.~Lukyanov for discussions and help.

\smallskip

\noindent This research is supported by DOE grant $\#$DE-FG02-96 ER 40949.

\smallskip

\noindent Research of AZ falls within the Federal Program
``Scientific and Scientific-Pedagogical Personnel of Innovational Russia''
on 2009-2013 (state contract No. 02.740.11.5165), and Russian Ministry
of Science and Technology under The Scientific Schools 6501.2010.2.

\bigskip
\bigskip


\end{document}